# Weathering the storm: How foreign aid and institutions affect entrepreneurship following natural disasters

**Christopher Boudreaux**
Florida Atlantic University

**Monica Escaleras**
Florida Atlantic University

**Anand Jha**
Wayne State University

## ABSTRACT

This study examines how foreign aid and institutions affect entrepreneurship activity following natural disasters. We use insights from the entrepreneurship, development, and institutions literature to develop a model of entrepreneurship activity in the aftermath of natural disasters. First, we hypothesize the effect of natural disasters on entrepreneurship activity depends on the amount of foreign aid received. Second, we hypothesize that natural disasters and foreign aid either encourages or discourages entrepreneurship activity depending on two important institutional conditions: the quality of government and economic freedom. The findings from our panel of 85 countries from 2006 to 2016 indicate that natural disasters are negatively associated with entrepreneurship activity, but both foreign aid and economic freedom attenuate this effect. In addition, we observe that foreign aid is positively associated with entrepreneurship activity but only in countries with high quality government. Hence, we conclude that the effect of natural disasters on entrepreneurship depends crucially on the quality of government, economic freedom, and foreign aid. Our findings provide new insights into how natural disasters and foreign aid affect entrepreneurship and highlight the important role of the institutional context.

# 1. Introduction

During the last two decades, the frequency of natural disasters has risen dramatically. Since 1998, there have been approximately 10,000 natural disasters worldwide that have killed 1.3 million people, affected more than 4 billion people, and caused over $2 trillion in estimated damages[1]. Moreover, the economic losses of natural disasters have been increasing over the last few decades, with the number of natural disasters causing substantial losses increasing by a factor of three since the 1980s (Hoeppe, 2016). Policymakers thus have an interest in identifying mechanisms that minimize the shock of natural disasters on society and the economy.

Recent research suggests entrepreneurship activity wanes following a natural disaster (Boudreaux, Escaleras, & Skidmore, 2019; Doern, Williams, & Vorley, 2019; Muñoz, Kimmitt, Kibler, & Farny, 2019), and several case studies provide additional support for this claim (Chamlee-Wright & Storr, 2009; Grube & Storr, 2018). A related line of research is the entrepreneurship literature on "crises" (Doern et al., 2019), which provides some understanding of how entrepreneurs can deal with unexpected adverse situations. This literature examines how entrepreneurs respond, prepare, and recover from natural disasters and other crises illustrating the importance of resilience (Bullough, Renko, & Myatt, 2014), a proper mindset and prior experience (Doern, 2016), social capital (Chamlee-Wright & Storr, 2009; Grube & Storr, 2018; Martinelli, Tagliazucchi, & Marchi, 2018), ambition and commitment (Davidsson & Gordon, 2016), and proper planning and access to resources (Runyan, 2006). Despite the importance of these findings, what is missing from the literature is an understanding of the policy mechanisms that might enable entrepreneurs to "bounce forward" (Muñoz et al., 2019) following a crisis event. While it is important to promote the implications from the crises literature—proper mindsets, commitment, preparedness, and resilience—it is also important to identify ways society can enable entrepreneurs to flourish in the aftermath of disasters. What can governments and key organizations do to aid and encourage entrepreneurs' recovery process? Hence understanding the mechanisms that alleviate the adverse impact of natural disaster on entrepreneurship remains an important question that has yet to be considered.

---

[1] https://news.un.org/en/story/2018/10/1022722

The purpose of this study is to examine several mechanisms that might mitigate the adverse effect of natural disasters on entrepreneurship activity. First, we examine whether foreign aid can accelerate the recovery process after a natural disaster strikes. At first glance, we might expect foreign aid to help attenuate the disruptive effects of disasters on entrepreneurship activity because it accelerates recovery and benefits society (Grube & Storr, 2018; Shepherd & Williams, 2014). However, findings from the corruption and development literature argue that foreign aid often does not meet its intended purpose because politicians and government officials are able to extract rents from the incoming monies (Easterly, 2002, 2003; Easterly & Pfutze, 2008; Leeson & Sobel, 2008). Second, based on the quality of government and public choice literature (Bailey & Thomas, 2017; Bertrand & Kramarz, 2002; Charron & Lapuente, 2013; Fukuyama, 2014; Rothstein, Charron, & Lapuente, 2013), we argue the efficacy of foreign aid ultimately depends on the quality of government and entrepreneurship critically depends on the ease of doing business and the extent of regulation in the economy (Bailey & Thomas, 2017; Djankov, La Porta, Lopez-de-Silanes, & Shleifer, 2002; Lucas & Boudreaux, 2020). In fact, Holcombe and Boudreaux (2015) find that large governments are not necessarily more corrupt. Rather, it is the regulatory state—not the productive or redistributive state—that is associated with more corruption (i.e., lower quality government). For these reasons, we disentangle the natural disaster and foreign aid effect in two additional hypotheses— (1) foreign aid has a positive effect on entrepreneurship activity as the quality of government increases, and (2) natural disasters have a less adverse effect on entrepreneurship activity as economic freedom increases.

To test our hypotheses, we merge natural disaster data from the *Centre for Research on the Epidemiology of Disasters* (CRED) with foreign aid from the *World Bank* and business start-up data from the *World Bank Group Enterprise Survey* database. Our sample consists of 85 countries over the period from 2006 to 2016. Using a fixed effects regression model, our evidence indicates that natural disasters indeed discourage entrepreneurship activity, but this disruptive effect on entrepreneurship activity is attenuated by the presence of foreign aid. In other words, in the absence of foreign aid, natural disasters discourage subsequent entrepreneurship activity more than when there is greater availability of foreign aid funds. Consistent with our predictions, we also find that foreign aid is associated with higher levels of entrepreneurship activity in countries with higher quality government, and greater economic freedom moderates the adverse effect of natural disasters on entrepreneurship activity.

Our study contributes to the entrepreneurship literature by documenting that besides personal traits like experience, ambition and resilience, assistance from foreign governments may alleviate some of the problems entrepreneurs face—particularly when the country receiving aid has a higher quality government. Similar to previous studies (Boudreaux, Escaleras, & Skidmore, 2019; Doern et al., 2019; Muñoz et al., 2019), we find that natural disasters can hamper entrepreneurship activity; however, we extend this line of research by suggesting that economic freedom attenuates this adverse effect. From the perspective of helping a country's entrepreneurs, our findings suggest that disaster struck governments should not shun foreign aid and should focus on improving the quality of government in society. We discuss the details of how our research extends some of the recent work in entrepreneurship in the discussion section.

## 2. Theory and Hypotheses
*2.1. The effect of natural disasters on entrepreneurship*

Theoretically, the way natural disasters affect entrepreneurship is unclear. On the one hand, natural disasters can generate entrepreneurial opportunities since destruction generates market inefficiencies. The ensuing opportunities can turn motivation into intention and action to provide new products and services (Monllor & Murphy, 2017). In addition, natural disasters can reduce the market entry requirement by reducing the opportunity cost of capital if other business opportunities become unprofitable. From this perspective, entrepreneurs often view "crises" as opportunities for new businesses and new collaborations at the individual, local and regional levels (Brück, Llussá, & Tavares, 2011; Linnenluecke & McKnight, 2017; Marino, Lohrke, Hill, Weaver, & Tambunan, 2008).

There are several case studies investigating the perceptions and actions taken by entrepreneurs following a natural disaster. These case studies often document an increase in entrepreneurial activity. Muñoz et al. (2019) found that entrepreneurs under continuous threat develop a level of preparedness that not only gives them the ability to re-build their business (bouncing back) but also look for new opportunities (bouncing forward). Also, Grube and Storr (2018) interview residents and business people after three natural disasters and offer evidence on how entrepreneurs contribute to the overall disaster recovery in three ways. First, entrepreneurs supply needed goods and services to victims. Second, entrepreneurs leverage social capital to navigate uncertainty, including questions about how they will clean up and rebuild. Lastly, entrepreneurs are motivated to rebuild due to a sense of attachment with a particular place.

On the other hand, however, natural disasters can hinder entrepreneurial activity by increasing entrepreneurial uncertainty and fear of failure (Monllor & Murphy, 2017). For example, natural disasters can damage infrastructure, disrupt supply chains, and affect firm performance (Altay & Ramirez, 2010; Carvalho, 2014). In addition, natural disasters can impede the return to normal operations (Chamlee-Wright & Storr, 2009; Grube & Storr, 2018) and thus decrease productivity (Boehm, Flaaen, & Pandalai-Nayar, 2019). For example, Runyan (2006) found that business owners faced many barriers to getting their businesses back "online" following Hurricane Katrina like a reduction in labor and access to capital, decrease in the number of customers, infrastructure disruptions (lack of electricity and phone service), damage to their buildings, and limited credit.

The majority of these studies are based on survey data and are qualitative by nature. There is only a few quantitative studies documenting the relationship between natural disasters and entrepreneurship. At the aggregate level, Klapper & Love (2011) use panel data for 93 countries and show that most countries experienced a sharp drop in new firm registration during the financial crisis of 2008. Similarly, Boudreaux, Escaleras, & Skidmore (2019) using cross country panel data show that natural disasters discourage economic development in the short-run by inhibiting new business start-up measured by the number of new firm registrations.

In this study we extend the work of these studies by showing empirically the negative impact of disasters on entrepreneurship activity. However, we go one step further by investigating the role of factors that can either attenuate or amplify the negative effect of natural disasters on entrepreneurship. We will discuss in further detail these factors and formulate our hypotheses in the following sections.

*2.2. Factors that moderate the relationship between natural disasters and entrepreneurship*

If the relationship between natural disasters and entrepreneurship is ambiguous, as the previous section suggests, it becomes important to identify potential moderating factors that can tip the scale in either direction. From a policy perspective, it is our goal to identify factors that can make natural disasters less burdensome on entrepreneurs. Several studies in the crises literature (Doern et al., 2019) offer clues toward such moderating factors.

Similar to natural disasters, the crises literature documents negative impacts of crises on entrepreneurship activity including business failure, contraction, and resource destruction (Doern, 2016). Importantly, these studies also identify factors that have the potential of moderating a crisis' negative impact on entrepreneurship. For example, Bullough et al. (2014), investigate entrepreneurial intention during the war in Afghanistan. They find that entrepreneurs' resilience reduces the adverse impact of perceived danger on entrepreneurial intentions. Martinelli et al. (2018) find that entrepreneurs' social capital enhances their resilience after a disaster. Doern (2016) finds that prior experience and a mindset to seek help can also increase resilience. Davidsson and Gordon (2016) find entrepreneurs with a higher level of commitment and ambition are less likely to disengage from attempting to create a business after a macro-economic crisis. Runyan (2006) suggests that the lack of planning and inability to access resources can impede an entrepreneur's recovery from a natural disaster.

In contrast to these studies focusing on entrepreneurs' resilience, experience and ambition to moderate the adverse impact of crises, we focus on the assistance from the international community—namely foreign aid as a moderating factor. We also adopt an institutional perspective that suggests the institutional context plays an important role in determining how natural disasters and foreign aid affect entrepreneurship activity.

*2.3. The moderating role of foreign aid on the effect of natural disasters on entrepreneurship*

Whether foreign aid in the aftermath of natural disasters attenuates or exacerbates the adverse effect of natural disasters is unclear. An increase in foreign aid funds can influence both the public sector and the private sector in the recipient country. For the public sector, foreign aid may enhance the accountability of political institutions (Eubank, 2012) and release governments from revenue constraints (Bräutigam & Knack, 2004). Thus, relaxing the recipient government's budget constraints may increase government investments and strengthen provisions of public goods and services, the latter of which relate positively to the flourishing of entrepreneurship activity (Audretsch, Heger, & Veith, 2015). In addition, foreign aid contributes to technology transfers (Sawada, Matsuda, & Kimura, 2012), facilitates building infrastructure (Miyamoto & Chiofalo, 2015), and promotes productive capacities (Lee et al., 2016). For example, Kligerman, Barry, Walmer, and Bendavid (2015) collected data before and after the 2010 Haiti earthquake and found that the period after the earthquake was associated with a higher number of healthcare and surgical facilities and inpatient beds. Foreign aid was a driving factor.[2] Hence, we hypothesize the following.

**Hypothesis 1a**: **Foreign aid attenuates the adverse effect of natural disasters on entrepreneurship activity**

The counterargument is as follows: post-disaster foreign aid often does not accomplish the intended purpose. Elites siphon off resources or manipulate aid's intended purpose for their own advantage (Easterly, 2002, 2003; Easterly & Pfutze, 2008; Leeson & Sobel, 2008).

Moreover, those in power can use foreign aid to further entrench their power and to answer their political needs (Dutta, Leeson, & Williamson, 2013; Dutta & Williamson, 2016). For

---

[2] 12 out of 13 new facilities that opened and seven out of eight facilities that were rebuilt were due to foreign aid.

example, after the 2004 tsunami in Sri Lanka, the age-old conflict affected the battle for resources and violence eventually erupted.³ Foreign aid could also result in an erosion of trust. When foreign organizations have an influential role in the disbursement of aid, it becomes easier for the government to blame foreigners for failures. Consequently, foreign aid makes it easier for the public officials to dodge their responsibility and accountability of domestic governments, hampering the speed of recovery. Furthermore, most foreign organizations want to bring in their own staff rather than hire from the local workforce. These foreign workers often have higher salaries than the market rate, and this hiring process can result in internal brain drain. Overall, the process stymies recovery efforts since local workers are not hired. This results in fewer entrepreneurs who are ready to use their business and entrepreneurial efforts to assist with the recovery process. Based on this reasoning, we state the following hypotheses:

**Hypothesis 1b: Foreign aid worsens the effect of natural disasters on entrepreneurship activity.**

*2.4. Does foreign aid's effect on entrepreneurship depend on the quality of government?*

William Baumol theorizes that entrepreneurial individuals can generate wealth for themselves in two ways: (i) join the private sector, or (ii) join efforts to manipulate the political and legal process. Their choices largely depend upon the rate of return for these two paths as well as the existing political and legal institutions (Baumol, 1990). Several studies support this theory—they all report a positive association between strong institutions and entrepreneurship (e.g., Aidis, Estrin, & Mickiewicz, 2008; Dutta, Sobel, & Roy, 2013; Estrin, Korosteleva, & Mickiewicz, 2013; Sobel, 2008).

Following a natural disaster and an influx of foreign aid, there will be opportunities. On the one hand, entrepreneurial individuals will have a chance to seize the opportunities that a natural disaster creates and foreign aid facilitates the start of a new business. On the other hand, they will have a chance to ally with politicians to siphon off foreign aid received.

---

³ http://theconversation.com/five-ways-foreign-aid-and-ngos-can-make-things-worse-when-disaster-strikes-50486

A country's institutional quality, particularly the quality of government measured by corruption (Charron & Lapuente, 2013; Rothstein et al., 2013), affects this relationship. In countries with lower quality of government, entrepreneurial individuals will find it attractive to participate in siphoning off foreign aid meant for victims of the natural disaster. In contrast, in countries with higher quality government there will be limited opportunities to steal foreign assistance and a higher incentive to exploit the opportunities disasters create to start a new business.

Another channel via which corruption might impede the positive effect of foreign aid is by reducing the amount of aid that reaches the people. A recent study by Andersen, Johannesen, and Rijkers (2020) finds that aid disbursements coincide with sharp increases in bank deposits in offshore financial centers that are well known for secrecy. Anecdotal evidence also supports this view. For example, in the Democratic Republic of Congo, about 85 percent of humanitarian aid does not reach the people (Riddell, 2008). Even in developed countries, where the problem of corruption is less severe, not all humanitarian aid reaches the intended people due to corrupt government officers. Riddell (2008) reports studies documenting that about $1.4 billion, 16 percent of the humanitarian government aid, was lost to fraud in the United States in the aftermath of Hurricanes Rita and Katrina. Regardless, misappropriation is more likely to occur in countries with poor institutions, where politicians feel less accountable to their citizens.

It is, therefore, possible that in countries with less corruption, the positive effect of foreign aid in mitigating the adverse impact of natural disasters will be stronger.

**Hypothesis 2: The positive effect of foreign aid on entrepreneurship activity is more pronounced when the aid recipient country has better control of corruption.**

*2.5. Does economic freedom attenuate the adverse effect of natural disasters?*

While low quality governments might render foreign aid ineffective, we also know the lack of economic freedom can discourage entrepreneurship (Aidis, Estrin, & Mickiewicz, 2012; Estrin et al., 2013). Although many countries have improved their economic freedom in the last couple of decades, the lack of economic freedom still appears to be a problem. Moulick, Pidduck, and Busenitz (2019) cites examples of how even in China, one of fastest growing emerging economies, obtaining a license to start some types of business can be difficult.

In countries with greater economic freedom the regulations are less cumbersome and property rights are much stronger. Boudreaux, Nikolaev, and Klein (2019) builds a theory of socio-cognitive traits and explain when a country has greater economic freedom an entrepreneur's self-efficacy—that is, the confidence in themselves, and alertness to new opportunities have a much stronger effect on starting a new business. Economic freedom also reduces the entrepreneurs' fear of failure. Lucas and Boudreaux (2020) uses U.S. data and finds that national regulation destroys jobs in states that have lower economic freedom partly because the lack of economic freedom impedes the establishment of new business. The key argument in these studies is that lack of economic freedom creates uncertainty and ambiguity on how to deal with unexpected situations and hence discourages opening of new business.

Building on these studies, we argue that in countries with greater economic freedom, entrepreneurs might be able to harness the opportunities that natural disasters create. This is because, following these events, fewer obstacles exist to starting a new business, there is lower fear of failure, and there is less ambiguity in dealing with unexpected situations in the future. Also, given the higher level of accountability that the citizens expect from their government in countries with higher economic freedom, entrepreneurs expect a speedy recovery following a disaster.

**Hypothesis 3: The adverse effect of natural disasters on entrepreneurship activity is attenuated by greater economic freedom.**

## 3. Data and Methods
*3.1. Data*

To test our hypotheses, we use panel data comprised of 85 countries over the period from 2006 to 2016. Because data are not available for all countries and years, the panel data are unbalanced, and the number of observations depends on the choice of explanatory variables. The time period and sample are due to merging variables and missing observations. Table 1 provides descriptive statistics for all variables, Table 2 provides the correlation matrix of the variables, and the appendix provides the definitions of the variables and their sources.

Our dependent variable, business start-up density (log), is the new business start-up rate, measured as the number of newly registered limited liability corporations per 1,000 working-age individuals (between 15 and 64 years old – i.e., new business density) and is taken from the World's Bank Doing Business database. It is important to note that limited liability implies that the liability of the owners of a firm is limited to their investment in the business. Due to differences and lack of consistency in the definition of partnerships and sole proprietorships across the world, those are not considered in the database. The dependent variable ranges from 0.002 to 11.88 with a mean of 1.49. We use the natural logarithm of this measure to ensure that our dependent variable is not skewed and satisfies ordinary least squares assumptions.

Our key explanatory variables include several measures of natural disasters' intensity and a measure of foreign aid. The data on natural disasters are taken from CRED at the Université

Catholique de Louvain, which compiles and publishes data on both natural and manmade disasters in their database, EM-DAT. Natural disasters include several different categories such as earthquakes, floods, volcanoes, slides, and windstorms (i.e., hurricanes and typhoons, tornados and cyclones, and other storms). CRED includes a "disaster" in their database if it results in 1) 10 or more people killed, 2) 100 or more people affected,[4] 3) the declaration of a state of emergency, or 4) a call for international assistance. We choose as the measure of *disaster intensity,* the sum of those affected, injured, or rendered homeless according to CRED. The disaster intensity variable ranges from 0 to 16.2 million with a mean of 417,917. Foreign aid is measured as the total Official Development Assistance (ODA) disbursements to a recipient country as a percentage of gross national investment (GNI) or as a per capita amount. This measure is taken from the World Bank's *Development Indicators*. This variable ranges from 0 to 51.42 and has a mean of 4.75.

Following the literature of the determinants for entrepreneurship, we include several control variables. First, we include three measures of new business entry regulation taken from the World Bank's *Doing Business database*: (i) the number of procedures that a start-up firm needs to go through to get legal status to operate as a firm, (ii) the time it takes to obtain legal status to operate as a firm (measured in business days), and (iii) the cost of obtaining legal status where these costs cover all identifiable expenses to obtain legal status. The World Bank defines procedures as any interaction between the founders of the new business and external parties necessary to complete the start-up process legally. The number of required procedures in our sample ranges between 2 and 19 and has a mean of 8.63. The days required to start a business captures the median duration that business incorporation lawyers deem necessary to complete the founding process. This measure ranges from 2 to 234 and has a mean of 27.91. Evidence shows the degree of regulation significantly diminishes entrepreneurial activity (Dreher & Gassebner, 2013; Dutta & Sobel, 2016; Klapper, Laeven, & Rajan, 2006). Hence, we include business start-up costs measured as a percentage of the respective country's income per capita, which ranges from 0.30 to 1,314.6 with a mean of 41.21.

---

[4] "*Affected*" is defined as "people requiring immediate assistance during a period of emergency, i.e. requiring basic survival needs such as food, water, shelter, sanitation and immediate medical assistance or the appearance of a significant number of cases of an infectious disease introduced in a region or a population that is usually free from disease."

  We also include socioeconomic variables that might influence start-up activity. The World Bank *Development Indicators* provide these data. First, we control for the potentially important effects of overall economic development, which could create more opportunities for new businesses. We include both the logarithm of the gross domestic product (GDP) per capita based on purchasing power parity and GDP growth (Audretsch & Keilbach, 2007). GDP per capita ranges from $773 to $45,636 USD, and GDP growth ranges from -52.43% to 21.02%. Second, to measure the level of financial development in a country, we include domestic credit provided by its financial sector (as a percentage of GDP). This variable ranges from -16.38% to 192.66% and has a mean of 48.4%. We expect that better access to financing leads to more opportunities to start a business (Klapper et al., 2006). We also control for population in the largest city (as a percentage of total population). This measure ranges from 2% to 50% and has a mean of 18.54%. Population density in cities generates positive externalities and opportunities for increasing returns, thus promoting entrepreneurial activity (Dutta & Sobel, 2016; Leeson & Sobel, 2008). Finally, we use a measure of corruption as a proxy for a country's quality of government. The *World Governance Indicators* (WGI) database provides this data, which captures the "perceptions of the extent to which public power is exercised for private gain, including both petty and grand forms of corruption, as well as capture of the state by elites and private interests" (Kaufmann, Kraay, & Mastruzzi, 2011). This variable ranges from -2.5 to 2.5 with higher numbers representing lower levels of corruption. For robustness, we also report results using a second measure of corruption from Transparency International's Corruption Perceptions Index (CPI). According to Dutta and Sobel (2016) and Boudreaux, Nikolaev, and Holcombe (2018), corruption has a negative impact on entrepreneurship activity. Finally, in additional analyses, we include data from the *Economic Freedom of the World Index* (EFW) to capture country-level institutions (Gwartney, Lawson, Hall, & Murphy, 2019)

**[Insert Tables 1 and 2 Here]**

*3.2. Methods*

We model the relationship between natural disasters and entrepreneurship activity using ordinary least squares (OLS) regression. We take advantage of the structure of our panel data and incorporate country and year fixed effects, which isolate the effect of natural disasters and foreign aid flows that occur *within* a country. Moreover, we use one-year lags of our key explanatory variables (i.e., natural disasters and foreign aid) to predict subsequent changes in business start-up rates in the following year. Based on the variables described in the previous section, we estimate the following model:

$$E_{it} = \beta_0 + \sum_{j=1}^{J} \alpha_j \ln D_{jit-1} + \gamma A_{it-1} + \sum_{j=1}^{J} \rho_j \left(D_{jit-1} \times A_{it-1}\right)$$
$$+ \sum_{k=1}^{K} \delta_k X_{kit} + \beta_2 v_i + \beta_3 \lambda_t + \varepsilon_{it} \qquad (1)$$

where $E_{it}$ is our measure of entrepreneurship measured by the new business density (start-up rate) for country *i* in year *t*. Because business start-up density is power law distributed (Crawford, Aguinis, Lichtenstein, Davidsson, & McKelvey, 2015; Crawford, McKelvey, & Lichtenstein, 2014), we log transform this measure. In additional robustness tests using System Generalized Method of Moments (GMM) estimation, we include a lag of the dependent variable to account for entrepreneurial dynamism, which is consistent with prior research on the topic (Dutta & Sobel, 2016). $D_{jit-1}$ is a matrix of four natural disaster variables (natural disaster intensity, number of persons affected, injured, and homeless). $D_{jit-1}$ denotes disaster $j$ in country $i$ in year $t$-1. $A_{it-1}$ denotes ODA in country $i$ in year $t$-1. ODA is our measure of foreign aid. We also include an interaction term between each of our four natural disaster variables and foreign aid. The parameter $\rho_{jit-1}$ captures these interactions. In additional tests, we separate the natural disaster variables into two categories: climatic (floods and windstorms) and geologic (earthquakes, slides, and volcanic eruptions). We transform these variables using the natural logarithm and add one to account for multiple zeros. We use disaster data from the previous year because contemporaneous values are unlikely to affect new business entry if the disaster occurs during the same year but in a later month as new businesses enter the market. $X_{kit}$ is the vector of $k$ control variables for each country $i$ in year $t$. $v_i$ is the country-specific fixed effect, $\lambda_t$ is the year fixed effect, and $\varepsilon_{it}$ is the error term. We use heteroscedastic-consistent standard errors that are robust clustered at the country level.

## 4. Results
### *4.1. Main results*

We report the results from our analysis of entrepreneurship activity in Table 3. Column 1 reports the baseline model that only includes control variables. Column 2 augments this model with a measure of natural disaster intensity: the number of persons affected, injured, and homeless due to natural disasters. Finally, Columns 3 through 6 separate the intensity measure by interacting each of its subcomponents with ODA. Our evidence suggests natural disasters discourage entrepreneurship activity, supporting previous findings (Boudreaux, Escaleras, and Skidmore, 2019). More importantly, our contribution is to show that foreign aid attenuates this relationship. Following Brambor, Clark, and Golder (2006) and Tchamyou and Asongu (2017), we report net effects which are the effect of interaction at the mean amount of ODA. For example, the corresponding net effect reported in Column 6 of Table 3 is -0.005 (– 0.011 + [4.75 x 0.0012]).[5]

**[Insert Table 3 Here]**

To ease the interpretation of our marginal effects, we report the interaction between foreign aid and natural disaster variables in Figure 1. According to this figure, the effect of natural disasters on entrepreneurship activity depends on how much foreign aid—if any—countries receive following a natural disaster. In the absence of receiving any foreign aid flows, natural disasters are associated with a slight decrease in entrepreneurship activity. In contrast, at the 75$^{th}$ percentile of foreign aid received (roughly one standard deviation above the mean), natural disasters are associated with a slight increase in entrepreneurship activity. Overall, we find a similar relationship for all our natural disaster variables. Thus, we conclude that the effect of natural disasters on entrepreneurship activity depends on the amount of foreign aid received.

**[Insert Figure 1 Here]**

*4.2. Does the Quality of Government Moderate the Relationship Between Foreign Aid and Entrepreneurship Activity?*

To test hypothesis 2—that the effect of foreign aid on entrepreneurship activity depends on a country's quality of government—we regress the measure of new business registrations on corruption, foreign aid (i.e., ODA), and their interaction. We report these results in columns 1 and 2 of Table 4 with the only difference being that we use the *World Governance Indicators* (WGI) measure of corruption in column 1, and we use *Transparency International's* (TI) measure of corruption in column 2. Overall, the findings suggest that foreign aid has a positive relationship with our measure of entrepreneurship activity, and this relationship becomes stronger as the quality of government increases. Note that higher numbers indicate less corruption for both indicators, which is consistent with the notion of "control of corruption" (Charron & Lapuente, 2013; Rothstein et al., 2013).

---

[5] The mean value of ODA is 4.75.

[Insert Table 4 Here]

To ease interpretation, we also report the moderating effect in Figure 2. The results suggest that foreign aid has little effect on entrepreneurship activity when countries have less control of corruption (i.e., a low score), and a positive effect on entrepreneurship activity when countries have more control of corruption (i.e., a high score). Although the results are stronger for the measure of corruption from *Transparency International*, the results are similar for both measures, lending additional support for hypothesis 2.

[Insert Figure 2 Here]

*4.3. Does Economic Freedom Moderate the Relationship Between Natural Disasters and Entrepreneurship Activity?*

To test hypothesis 3—the adverse effect of natural disasters on entrepreneurship activity is attenuated by greater economic freedom—we regress new business registrations, our measure of entrepreneurship activity, on disaster intensity, economic freedom, and their interaction. We also include all control variables listed earlier. We use the economic freedom index constructed by Gwartney et al. (2019). The economic freedom index measures freedom to decide what to be produced and consumed, freedom to keep what you earn, freedom to exchange with foreigners, and protection of money as a store of value. Research shows that institutional variables associated with economic freedom are most strongly associated with entrepreneurship (Nikolaev, Boudreaux, & Palich, 2018). **Following Lucas and Boudreaux (2020), we also report our results using the measure of economic freedom but omitting area 1—the size of government indicator—from** **the index, since findings suggests government size and the overall economic freedom index is uncorrelated with the rest of the index (Aidis et al., 2012; Bergh, 2020). In this index, countries with a large government sector relative to the private sector receive a low score in area 1. As a result, lower numbers indicate smaller size governments.**

The results reported in columns 3 and 4 of Table 4 reveal that economic freedom (EFW) is associated with more entrepreneurship activity, and natural disasters are associated with less entrepreneurship activity. Importantly, we observe that increases in economic freedom attenuate the negative effect of natural disasters on entrepreneurship activity. That is, the results suggest natural disasters have a less adverse effect on entrepreneurship activity as economic freedom increases.

To ease interpretation, we also report this moderating effect in Figure 3. The results indicate that natural disasters typically have a negative effect on entrepreneurship activity, which is consistent with the results found by Boudreaux, Escaleras, and Skidmore (2019). However, the results also suggest that economic freedom positively moderates this relationship. That is, natural disasters have negative effects on entrepreneurship activity at low levels of economic freedom but no effect or even a positive effect at higher levels of economic freedom. These findings support hypothesis 3.

**[Insert Figure 3 Here]**

*4.4. Supplemental Analysis*

We tested the sensitivity of our findings using models such as System GMM and instrumental variables. We also examined whether our findings differ by disaster type (climatic vs geologic) or the lag structure. Overall, our results are qualitatively similar using these various robustness checks, which are available in our supplemental appendix.

## 5. Discussion and conclusion

Our findings suggest natural disasters impact on entrepreneurship depends on foreign aid. Perhaps more importantly, our study identifies the institutional context as playing an important moderating role for the effect of natural disasters and foreign aid on entrepreneurship activity. Specifically, in the absence of corruption, foreign aid has a positive effect on entrepreneurship activity. In addition, the effect of natural disasters on the rate of entrepreneurship is less detrimental or even positive when countries have high levels of economic freedom. We contribute to both the entrepreneurship and development literature as well as institutional theory.

*5.1. Contributions to the Entrepreneurship Literature*

Research shows that because of the trauma they inflict, crises such as natural disasters make individuals risk-averse (Cassar, Healy, & Von Kessler, 2017). Individuals considering starting a new business are less eager to do so (Boudreaux, Escaleras, & Skidmore, 2019) as the perceptions of misfortune increase (Shahriar & Sepherd, 2019), and disasters damage the infrastructure conducive to starting a business.

Although studies have investigated factors that might alleviate some of the problems a crisis creates for entrepreneurs, most of these studies have adopted a behavioral approach and conducted qualitative rather than quantitative studies. They point out that entrepreneurs that can improvise, who are proactive, persistent, and determined fare better (e.g., Doern, 2016; Doern et al., 2019; Linnenluecke & Griffiths, 2010; Runyan, 2006; Williams, Gruber, Sutcliffe, Shepherd, & Zhao, 2017). However, there are only a few empirical studies. These studies rely heavily on survey responses from entrepreneurs and confirm that personal traits like resilience matters on the entrepreneurial perceptions and intentions (Bullough et al., 2014; Monllor & Murphy, 2017). Sometimes the results have been surprising. Davidsson and Gordon (2016) examine whether a macroeconomic crisis triggers behavioral responses such as disengagement, delay, and adaption, and whether certain personal traits can minimize this effect. They find that commitment and ambition reduce the likelihood of disengagement. Other than these two hypotheses, none of their eleven hypotheses are supported. Davidsson and Gordon (2016) conclude that their "most important finding is the relative *absence* of behavioral crisis responses" (p. 915). They argue future research should focus on "environmental jolts" which might exert a stronger effect on nascent entrepreneurs' behavior. They do not highlight the importance of non-behavioral influences.

We extend this line of research by suggesting that besides personal traits, institutional factors might also alleviate the problems created by a crisis. The quality of government and economic freedom might create a better environment for entrepreneurs. More economic freedom helps alleviate some of the adverse effects of the trauma a natural disaster caused by accelerating the recovery process and facilitating new infrastructure and the transfer of new technology, thus ultimately fostering entrepreneurship. The quality of government (Charron & Lapuente, 2013; Rothstein et al., 2013) also influences the recovery process. While our findings suggest foreign aid can play a pivotal role in aiding recovery and assisting entrepreneurship activity following a natural disaster, our study finds that this mechanism depends critically on the quality of government. In corrupt environments, foreign aid has little effect on entrepreneurship activity, but in environments with low corruption, foreign aid has a positive effect on entrepreneurship activity. Our findings thus make an important contribution to the literature. Specifically, our findings directly address calls in the management and entrepreneurship literature to better understand mechanisms to alleviate the adverse consequences of natural disasters on businesses and entrepreneurs (Van Der Vegt, Essens, Wahlström, & George, 2015).

*5.2. Contributions to the development and foreign aid literature*

Many researchers are skeptical of whether foreign-aid is beneficial for recipient country's citizens (De Mesquita & Smith, 2007, 2009; Easterly, Levine, & Roodman, 2004). Some argue it can instead concentrate the elite's power at the expense of the citizenry (De Mesquita & Smith, 2010). This power can empower the elite to maintain the status-quo, and resist creative destruction and thus thwart the implementation of new ideas (Acemoglu & Robinson, 2012).

Others suggest the impact of foreign aid on institutions and economic growth might be more complicated. For example, Dutta and Williamson (2016) find that foreign aid may improve free press in democratic countries, but not in autocratic countries. Dutta, Leeson, et al. (2013) propose that foreign aid might not improve the political institutions, but rather amplify them. They find that the average polity score—an index that measures the extent of democracy—is amplified in countries that receive foreign aid. There are many other studies that argue that relationships between foreign aid, economic growth and institutions is much more nuanced (e.g., Morrison, 2007; Wright, 2009; Yuichi Kono & Montinola, 2009).

Still, others suggest that foreign aid is often politically motivated and hence unsuccessful in spurring economic growth. For example, we know that "the direction of foreign aid is dictated as much by political and strategic considerations, as by the economic needs and policy performance of the recipients" (Alesina & Dollar, 2000, p. 33). Even humanitarian aid seems to be politically motivated. Drury, Olson, and Van Belle (2005) examine humanitarian assistance by the US and note that the "initial decision to grant [humanitarian] aid is the most political, but the subsequent dollar amount decision is also not devoid of political considerations" (p. 455).

By documenting that regardless of the intention behind foreign aid, it can have a positive effect on entrepreneurship after a natural disaster, our findings complement the view of Easterly (2003), who says that foreign aid has mostly no positive impact, but can be helpful under certain circumstances. In this respect, we provide a fresh dimension to the foreign aid literature. If foreign aid indeed fosters entrepreneurship in the first few years after a natural disaster, as we find, our results show that although foreign aid may not be useful in developing better institutions (Asongu, 2015; Asongu & Nwachukwu, 2017; Heckelman & Knack, 2009), it can be useful in encouraging entrepreneurs, and hence reduce income inequality—which may be a much more effective way to bring change in developing economies (Asongu, 2016)

*5.3. Policy Implications*

There is growing evidence that creating a conducive atmosphere for entrepreneurs in the wake of a natural disaster will benefit society (Grube & Storr, 2018). In fact, a key takeaway from the literature is that individuals undertake entrepreneurial activity that seeks to address current challenges, limiting the negative impacts of crises allowing them to be overcome (Grube & Storr, 2018). Shepherd and Williams (2014) note that "Ventures created in the aftermath of a natural disaster, given local knowledge and unencumbered by pre-existing systems, procedures, and capabilities, are highly effective at connecting the broader community with the local community through customizing resources to meet victims' needs and to quickly delivering these resources to alleviate suffering" (p. 952). The same authors illustrate that following the 2010 Haiti earthquake new venture creation not only reduced suffering but also facilitated the resilience of community members (Williams and Shepherd, 2016).

Our study suggests that soliciting foreign aid in the aftermath of a disaster will alleviate the adverse impact of disasters by encouraging entrepreneurship activity, particularly in the form of ventures addressing the challenges associated with the aftermath of disasters. This finding is a particularly useful insight in light of several studies that uncover the negative effects of foreign aid (Bräutigam & Knack, 2004; Rajan & Subramanian, 2007) and the crowding out of private contributions.

Policymakers also often argue that an environment conducive to start-ups' entry and survival is connected to peace and stability (Koltai, 2016). They argue that more foreign aid should be devoted to entrepreneurship.[6] The results of our study suggest that even when money may not be directly allocated for start-ups, foreign aid after a natural disaster can help entrepreneurship. In other words, if advocates cannot have donors allocate money to entrepreneurship directly, they could advocate for more aid in the aftermath of a natural disaster, which would foster entrepreneurship. Our study suggests that the policymakers of countries facing a natural disaster should be more welcoming of foreign aid in the aftermath of natural disasters.

*5.4. Implication on our understanding of the COVID-19 pandemic and entrepreneurship*

**Similar to a natural disaster, the COVID-19 pandemic is also a crisis. But, what is unique is its massive scale worldwide. Research shows small businesses are particularly hit hard by this pandemic (Fairlie, 2020). For those in developing countries that rely on foreign aid, the impact is even worse, as their governments lack resources and are ill-prepared to help small businesses (Bottan, Hoffmann, & Vera-Cossio, 2020)[7]. Several studies investigate how we can alleviate some of their concerns. For example, Akpan, Soopramanien, and Kwak (2020) argue that cutting-edge technology can help small businesses navigate the challenges they face during the pandemic by improvising and adopting new business models (Vaccaro, Getz, Cohen, Cole, & Donnally III, 2020). Our study contributes to this stream of literature by suggesting that providing foreign aid—especially if the corruption is lower in a country—**

---

[6] https://hbr.org/2016/08/entrepreneurship-needs-to-be-a-bigger-part-of-us-foreign-aid
**[7] https://www.intracen.org/news/Op-ed-The-effects-of-the-Covid-19-pandemic-on-small-businesses-and-global-value-chains/**

**can alleviate some of the adverse impacts of COVID-19 on entrepreneurs. This finding is particularly sobering because developed countries are themselves in a recession (Dabla-Norris, Minoiu, & Zanna, 2015) and foreign aid will likely decline (Cardwell & Ghazalian, 2020). Our findings suggest that this pandemic might be an opportunity for entrepreneurs (Shepherd, 2020) that foreign aid might help foster.**

*5.5.Limitations and future research*

As with any study, ours has a few limitations. For one, our data capture the amount of a country's official development assistance inflows. However, our data do not indicate what share of that assistance is attributed to the recovery following a natural disaster. Fortunately, our study incorporates country and year fixed effects, which helps to attenuate this problem. By analyzing changes *within a country* from one year to the next, we can observe changes in official development assistance inflows **reflected as demeaned variables.** This design approximates the share of assistance attributed to natural disasters. Future research could complement ours by investigating how the amount of aid allocated to natural disasters affects entrepreneurship activity. Also, because we do not have data on the types of start-ups, we are unable to compare the effect of natural disasters across industries.

Our study also raises a few related questions. We focus on entrepreneurship activity measured by new business registrations. A related but separate outcome worthy of investigation is to measure the survival of existing businesses following a disaster and see if foreign aid moderates the effect of natural disasters on business survival. Another extension of our study is to analyze the variation in the effect of natural disasters on new businesses in different industries. While wholesale and retail businesses generally report significant sales losses, manufacturing and construction often show gains following a disaster (Webb, Tierney, & Dahlhamer, 2000). Disasters can cause consumer preferences to change and thus influence the market demand for some products and services. Specifically, households will put more of their expenditures into reconstructing their homes and replacing damaged furnishings after a natural disaster. Consequently, disaster-relevant industries such as construction, building materials, and home and office furnishing can experience an increase in demand from disaster-stricken communities.

Another interesting investigation is to compare the impact of natural disasters on small, medium, and large businesses after a natural disaster strikes. Large firms are more likely to spread their risk by operating in multiple locations. In addition, large businesses are more likely to be able to afford hazard insurance and have contingency funds for disaster recovery. Furthermore, large firms might have financial and political influence in their communities, which gives them high priority in governmental recovery programs as well as substantial influence so that private contractors rebuild their facilities first. Consequently, small businesses experience more obstacles than larger firms in recovering from natural disasters.

Finally, while our study suggests that economic freedom alleviates the problems that a natural disaster creates, it does not tell which aspect of institutions matter the most. Another extension of this study would be to focus on examining which types of institutions matter.

**References**

Acemoglu, D., & Robinson, J. A. (2012). *Why nations fail: The origins of power, prosperity, and poverty*: Currency.
Aidis, R., Estrin, S., & Mickiewicz, T. (2008). Institutions and entrepreneurship development in Russia: A comparative perspective. *Journal of Business Venturing, 23*(6), 656-672.
Aidis, R., Estrin, S., & Mickiewicz, T. M. (2012). Size matters: entrepreneurial entry and government. *Small Business Economics, 39*(1), 119-139.
Akpan, I. J., Soopramanien, D., & Kwak, D.-H. (2020). Cutting-edge technologies for small business and innovation in the era of COVID-19 global health pandemic. *Journal of Small Business & Entrepreneurship*, 1-11.
Alesina, A., & Dollar, D. (2000). Who gives foreign aid to whom and why? *Journal of Economic Growth, 5*(1), 33-63.
Altay, N., & Ramirez, A. (2010). Impact of disasters on firms in different sectors: implications for supply chains. *Journal of Supply Chain Management, 46*(4), 59-80.
Andersen, J. J., Johannesen, N., & Rijkers, B. (2020). Elite Capture of Foreign Aid: Evidence from Offshore Bank Accounts. *World Bank Policy Research Working Paper*(9150).
Asongu, S. (2016). Reinventing foreign aid for inclusive and sustainable development: Kuznets, Piketty and the great policy reversal. *Journal of Economic Surveys, 30*(4), 736-755.
Asongu, S. A. (2015). The evolving debate of the effect of foreign aid on corruption and institutions in Africa. In *Handbook on the Economics of Foreign Aid*: Edward Elgar Publishing.
Asongu, S. A., & Nwachukwu, J. C. (2017). Is the threat of foreign aid withdrawal an effective deterrent to political oppression? Evidence from 53 African countries. *Journal of Economic Issues, 51*(1), 201-221.
Audretsch, D. B., Heger, D., & Veith, T. (2015). Infrastructure and entrepreneurship. *Small Business Economics, 44*(2), 219-230.
Audretsch, D. B., & Keilbach, M. (2007). The theory of knowledge spillover entrepreneurship. *Journal of Management Studies, 44*(7), 1242-1254.
Bailey, J. B., & Thomas, D. W. (2017). Regulating away competition: The effect of regulation on entrepreneurship and employment. *Journal of Regulatory Economics, 52*(3), 237-254.
Baumol, W. J. (1990). Entrepreneurship: Productive, unproductive, and destructive. *Journal of Political Economy, 98*(5), 893-921.
Bergh, A. (2020). Hayekian welfare states: explaining the coexistence of economic freedom and big government. *Journal of Institutional Economics, 16*(1), 1-12.
Bertrand, M., & Kramarz, F. (2002). Does entry regulation hinder job creation? Evidence from the French retail industry. *Quarterly Journal of Economics, 117*(4), 1369-1413.
Boehm, C. E., Flaaen, A., & Pandalai-Nayar, N. (2019). Input linkages and the transmission of shocks: Firm-level evidence from the 2011 Tōhoku earthquake. *Review of Economics and Statistics, 101*(1), 60-75.
Bottan, N., Hoffmann, B., & Vera-Cossio, D. (2020). The unequal impact of the coronavirus pandemic: Evidence from seventeen developing countries. *PloS one, 15*(10), e0239797.
Boudreaux, C. J., Escaleras, M. P., & Skidmore, M. (2019). Natural disasters and entrepreneurship activity. *Economics Letters*(182), 82-85.

Boudreaux, C. J., Nikolaev, B. N., & Holcombe, R. G. (2018). Corruption and destructive entrepreneurship. *Small Business Economics, 51*(1), 181-202.

Boudreaux, C. J., Nikolaev, B. N., & Klein, P. (2019). Socio-cognitive traits and entrepreneurship: The moderating role of economic institutions. *Journal of Business Venturing, 34*(1), 178-196.

Brambor, T., Clark, W. R., & Golder, M. (2006). Understanding interaction models: Improving empirical analyses. *Political Analysis*, 63-82.

Bräutigam, D. A., & Knack, S. (2004). Foreign aid, institutions, and governance in sub-Saharan Africa. *Economic Development and Cultural Change, 52*(2), 255-285.

Brück, T., Llussá, F., & Tavares, J. A. (2011). Entrepreneurship: The role of extreme events. *European Journal of Political Economy, 27*, 78-88.

Bullough, A., Renko, M., & Myatt, T. (2014). Danger zone entrepreneurs: the importance of resilience and self–efficacy for entrepreneurial intentions. *Entrepreneurship Theory and Practice, 38*(3), 473-499.

Cardwell, R., & Ghazalian, P. L. (2020). COVID-19 and International Food Assistance: Policy proposals to keep food flowing. *World Development, 135*, 105059.

Carvalho, V. M. (2014). From micro to macro via production networks. *Journal of Economic Perspectives, 28*(4), 23-48.

Cassar, A., Healy, A., & Von Kessler, C. (2017). Trust, risk, and time preferences after a natural disaster: experimental evidence from Thailand. *World Development, 94*, 90-105.

Chamlee-Wright, E., & Storr, V. H. (2009). "There's no place like New Orleans": sense of place and community recovery in the Ninth Ward after Hurricane Katrina. *Journal of Urban Affairs, 31*(5), 615-634.

Charron, N., & Lapuente, V. (2013). Why do some regions in Europe have a higher quality of government? *Journal of Politics, 75*(3), 567-582.

Crawford, C., Aguinis, H., Lichtenstein, B., Davidsson, P., & McKelvey, B. (2015). Power law distributions in entrepreneurship: Implications for theory and research. *Journal of Business Venturing, 30*(5), 696-713.

Crawford, C., McKelvey, B., & Lichtenstein, B. B. (2014). The empirical reality of entrepreneurship: How power law distributed outcomes call for new theory and method. *Journal of Business Venturing Insights, 1*, 3-7.

Dabla-Norris, E., Minoiu, C., & Zanna, L.-F. (2015). Business cycle fluctuations, large macroeconomic shocks, and development aid. *World Development, 69*, 44-61.

Davidsson, P., & Gordon, S. R. (2016). Much ado about nothing? The surprising persistence of nascent entrepreneurs through macroeconomic crisis. *Entrepreneurship Theory and Practice, 40*(4), 915-941.

De Mesquita, B. B., & Smith, A. (2007). Foreign aid and policy concessions. *Journal of Conflict Resolution, 51*(2), 251-284.

De Mesquita, B. B., & Smith, A. (2009). A political economy of aid. *International Organization, 63*(2), 309-340.

De Mesquita, B. B., & Smith, A. (2010). Leader survival, revolutions, and the nature of government finance. *American Journal of Political Science, 54*(4), 936-950.

Djankov, S., La Porta, R., Lopez-de-Silanes, F., & Shleifer, A. (2002). The regulation of entry. *The Quarterly Journal of Economics, 117*(1), 1-37.

Doern, R. (2016). Entrepreneurship and crisis management: The experiences of small businesses during the London 2011 riots. *International Small Business Journal, 34*(3), 276-302.

Doern, R., Williams, N., & Vorley, T. (2019). Special issue on entrepreneurship and crises: business as usual? An introduction and review of the literature. *Entrepreneurship & Regional Development, 31*(5-6), 400-412.

Dreher, A., & Gassebner, M. (2013). Greasing the wheels? The impact of regulations and corruption on firm entry. *Public Choice, 155*(3-4), 413-432.

Drury, A. C., Olson, R. S., & Van Belle, D. A. (2005). The politics of humanitarian aid: US foreign disaster assistance, 1964–1995. *The Journal of Politics, 67*(2), 454-473.

Dutta, N., Leeson, P. T., & Williamson, C. R. (2013). The amplification effect: foreign aid's impact on political institutions. *Kyklos, 66*(2), 208-228.

Dutta, N., & Sobel, R. (2016). Does corruption ever help entrepreneurship? *Small Business Economics, 47*(1), 179-199.

Dutta, N., Sobel, R. S., & Roy, S. (2013). Entrepreneurship and political risk. *Journal of Entrepreneurship and Public Policy, 2*(2), 130 - 143.

Dutta, N., & Williamson, C. R. (2016). Can foreign aid free the press? *Journal of Institutional Economics, 12*(3), 603-621.

Easterly, W. (2002). The cartel of good intentions: the problem of bureaucracy in foreign aid. *The Journal of Policy Reform, 5*(4), 223-250.

Easterly, W. (2003). Can foreign aid buy growth? *Journal of Economic Perspectives, 17*(3), 23-48.

Easterly, W., Levine, R., & Roodman, D. (2004). Aid, policies, and growth: comment. *American Economic Review, 94*(3), 774-780.

Easterly, W., & Pfutze, T. (2008). Where does the money go? Best and worst practices in foreign aid. *Journal of Economic Perspectives, 22*(2), 29-52.

Estrin, S., Korosteleva, J., & Mickiewicz, T. (2013). Which institutions encourage entrepreneurial growth aspirations? *Journal of Business Venturing, 28*(4), 564-580.

Eubank, N. (2012). Taxation, political accountability and foreign aid: Lessons from Somaliland. *Journal of Development Studies, 48*(4), 465-480.

Fairlie, R. W. (2020). *The impact of Covid-19 on small business owners: Evidence of early-stage losses from the April 2020 current population survey* (No. 0898-2937): National Bureau of Economic Researcho. Document Number)

Fukuyama, F. (2014). *Political order and political decay: From the industrial revolution to the globalization of democracy*: Macmillan.

Grube, L. E., & Storr, V. H. (2018). Embedded entrepreneurs and post-disaster community recovery. *Entrepreneurship & Regional Development, 30*(7-8), 800-821.

Gwartney, J., Lawson, R., Hall, J., & Murphy, R. (2019). *Economic freedom of the world: 2019 annual report*: The Fraser Institute.

Heckelman, J. C., & Knack, S. (2009). Aid, economic freedom, and growth. *Contemporary Economic Policy, 27*(1), 46-53.

Hoeppe, P. (2016). Trends in weather related disasters–Consequences for insurers and society. *Weather and Climate Extremes, 11*, 70-79.

Holcombe, R. G., & Boudreaux, C. J. (2015). Regulation and corruption. *Public Choice, 164*(1-2), 75-85.

Kaufmann, D., Kraay, A., & Mastruzzi, M. (2011). The worldwide governance indicators: methodology and analytical issues. *Hague Journal on the Rule of Law, 3*(2), 220-246.

Klapper, L., Laeven, L., & Rajan, R. (2006). Entry regulation as a barrier to entrepreneurship. *Journal of Financial Economics, 82*(3), 591-629.

Klapper, L., & Love, I. (2011). The impact of the financial crisis on new firm registration. *Economics Letters, 113*(1), 1-4.

Kligerman, M., Barry, M., Walmer, D., & Bendavid, E. (2015). International aid and natural disasters: a pre-and post-earthquake longitudinal study of the healthcare infrastructure in Leogane, Haiti. *The American Journal of Tropical Medicine and Hygiene, 92*(2), 448-453.

Koltai, S. R. (2016). *Peace Through Entrepreneurship: Investing in a Startup Culture for Security and Development*: Brookings Institution Press.

Lee, B. X., Kjaerulf, F., Turner, S., Cohen, L., Donnelly, P. D., Muggah, R., et al. (2016). Transforming our world: implementing the 2030 agenda through sustainable development goal indicators. *Journal of public health policy, 37*(1), 13-31.

Leeson, P. T., & Sobel, R. S. (2008). Weathering corruption. *The Journal of Law and Economics, 51*(4), 667-681.

Linnenluecke, M., & Griffiths, A. (2010). Beyond adaptation: resilience for business in light of climate change and weather extremes. *Business & Society, 49*(3), 477-511.

Linnenluecke, M. K., & McKnight, B. (2017). Community resilience to natural disasters: the role of disaster entrepreneurship. *Journal of Enterprising Communities: People and Places in the Global Economy, 11*(1), 166-185.

Lucas, D. S., & Boudreaux, C. J. (2020). National regulation, state-level policy, and local job creation in the United States: A multilevel perspective. *Research Policy, 49*(4), 1039-1052.

Marino, L. D., Lohrke, F. T., Hill, J. S., Weaver, K. M., & Tambunan, T. (2008). Environmental shocks and SME alliance formation intentions in an emerging economy: Evidence from the Asian financial crisis in Indonesia. *Entrepreneurship Theory and Practice, 32*(1), 157-183.

Martinelli, E., Tagliazucchi, G., & Marchi, G. (2018). The resilient retail entrepreneur: dynamic capabilities for facing natural disasters. *International Journal of Entrepreneurial Behavior & Research, 24*(7), 1222-1243.

Miyamoto, K., & Chiofalo, E. (2015). *Official development finance for infrastructure: Support by multilateral and bilateral development partners* (Vol. 25): OECD Publishing Paris.

Monllor, J., & Murphy, P. J. (2017). Natural disasters, entrepreneurship, and creation after destruction: A conceptual approach. *International Journal of Entrepreneurial Behavior & Research, 23*(4), 618-637.

Morrison, K. M. (2007). Natural resources, aid, and democratization: A best-case scenario. *Public Choice, 131*(3-4), 365-386.

Moulick, A. G., Pidduck, R. J., & Busenitz, L. W. (2019). Bloom where planted: Entrepreneurial catalyzers amidst weak institutions. *Journal of Business Venturing Insights, 11*, e00127.

Muñoz, P., Kimmitt, J., Kibler, E., & Farny, S. (2019). Living on the slopes: entrepreneurial preparedness in a context under continuous threat. *Entrepreneurship & Regional Development, 31*(5-6), 413-434.

Nikolaev, B. N., Boudreaux, C. J., & Palich, L. (2018). Cross-country determinants of early-stage necessity and opportunity-motivated entrepreneurship: accounting for model uncertainty. *Journal of Small Business Management, 56*, 243-280.

Rajan, R., & Subramanian, A. (2007). Does aid affect governance? *American Economic Review, 97*(2), 322-327.

Riddell, R. C. (2008). *Does foreign aid really work?* : Oxford University Press.
30

Rothstein, B., Charron, N., & Lapuente, V. (2013). *Quality of government and corruption from a European perspective: a comparative study on the quality of government in EU regions*: Edward Elgar Publishing.

Runyan, R. C. (2006). Small business in the face of crisis: identifying barriers to recovery from a natural disaster 1. *Journal of Contingencies and Crisis Management, 14*(1), 12-26.

Sawada, Y., Matsuda, A., & Kimura, H. (2012). On the role of technical cooperation in international technology transfers. *Journal of International Development, 24*(3), 316-340.

Shahriar, A. Z. M., & Shepherd, D. A. (2019). Violence against women and new venture initiation with microcredit: Self-efficacy, fear of failure, and disaster experiences. *Journal of Business Venturing, 34*(6), 105945.

Shepherd, D. A. (2020). COVID 19 and entrepreneurship: Time to pivot? *Journal of Management Studies, 57*(8), 1750-1753.

Shepherd, D. A., & Williams, T. A. (2014). Local venturing as compassion organizing in the aftermath of a natural disaster: The role of localness and community in reducing suffering. *Journal of Management Studies, 51*(6), 952-994.

Sobel, R. S. (2008). Testing Baumol: Institutional quality and the productivity of entrepreneurship. *Journal of Business Venturing, 23*(6), 641-655.

Tchamyou, V. S., & Asongu, S. A. (2017). Information sharing and financial sector development in Africa. *Journal of African Business, 18*(1), 24-49.

Vaccaro, A. R., Getz, C. L., Cohen, B. E., Cole, B. J., & Donnally III, C. J. (2020). Practice management during the COVID-19 pandemic. *The Journal of the American Academy of Orthopaedic Surgeons, 28*(11), 464-470.

Van Der Vegt, G. S., Essens, P., Wahlström, M., & George, G. (2015). Managing risk and resilience. *Academy of Management, 58*(4), 971-980.

Webb, G. R., Tierney, K. J., & Dahlhamer, J. M. (2000). Businesses and disasters: Empirical patterns and unanswered questions. *Natural Hazards Review, 1*(2), 83-90.

Williams, T. A., Gruber, D. A., Sutcliffe, K. M., Shepherd, D. A., & Zhao, E. Y. (2017). Organizational response to adversity: Fusing crisis management and resilience research streams. *Academy of Management Annals, 11*(2), 733-769.

Williams, T. A., & Shepherd, D. A. (2016). Building resilience or providing sustenance: Different paths of emergent ventures in the aftermath of the Haiti earthquake. *Academy of Management Journal, 59*(6), 2069-2102.

Wright, J. (2009). How foreign aid can foster democratization in authoritarian regimes. *American Journal of Political Science, 53*(3), 552-571.

Yuichi Kono, D., & Montinola, G. R. (2009). Does foreign aid support autocrats, democrats, or both? *The Journal of Politics, 71*(2), 704-718.

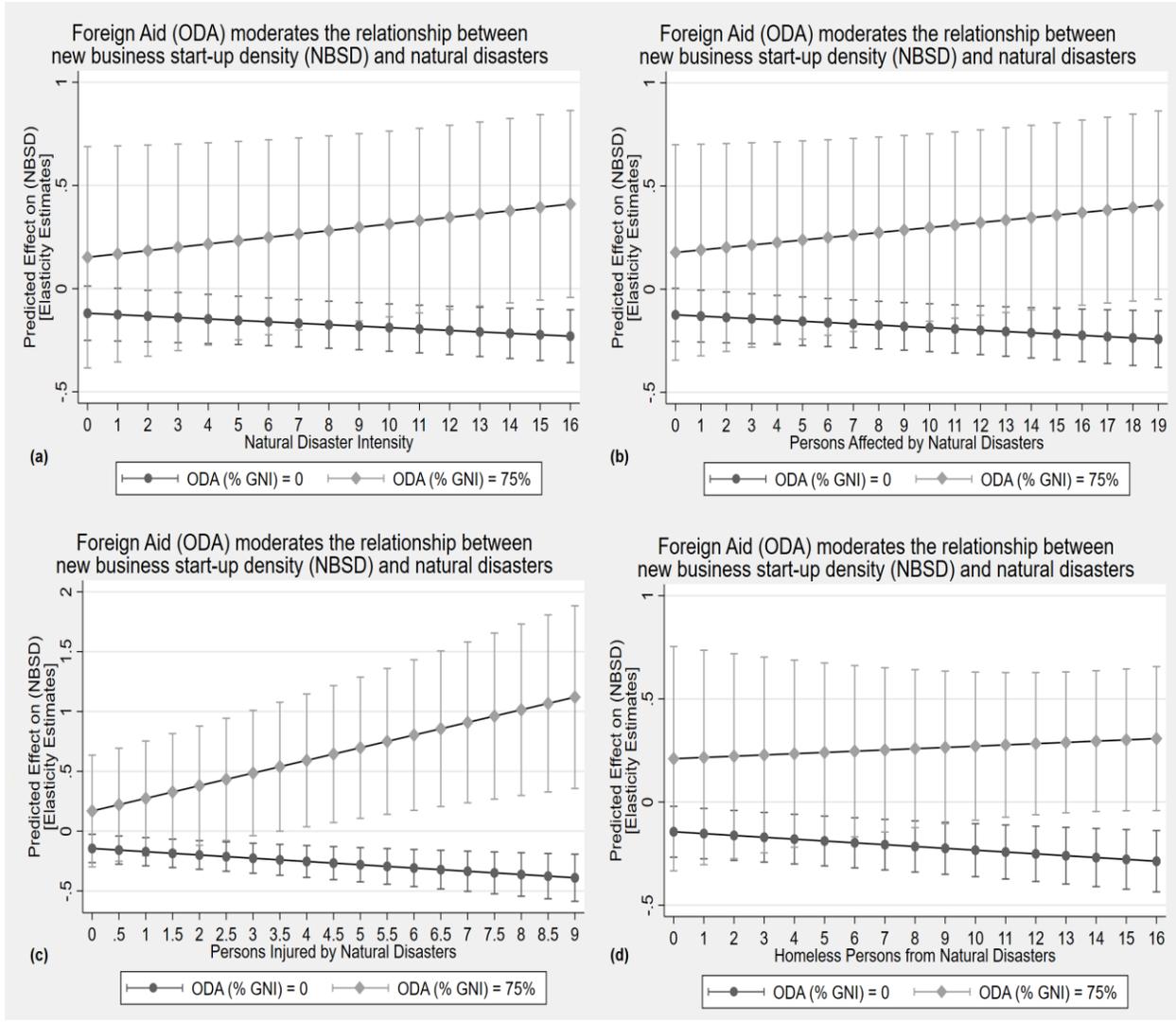

**Figure 1.** Moderating Effect of Foreign Aid (% GNI) on Natural Disasters (log) and Start-Up Density

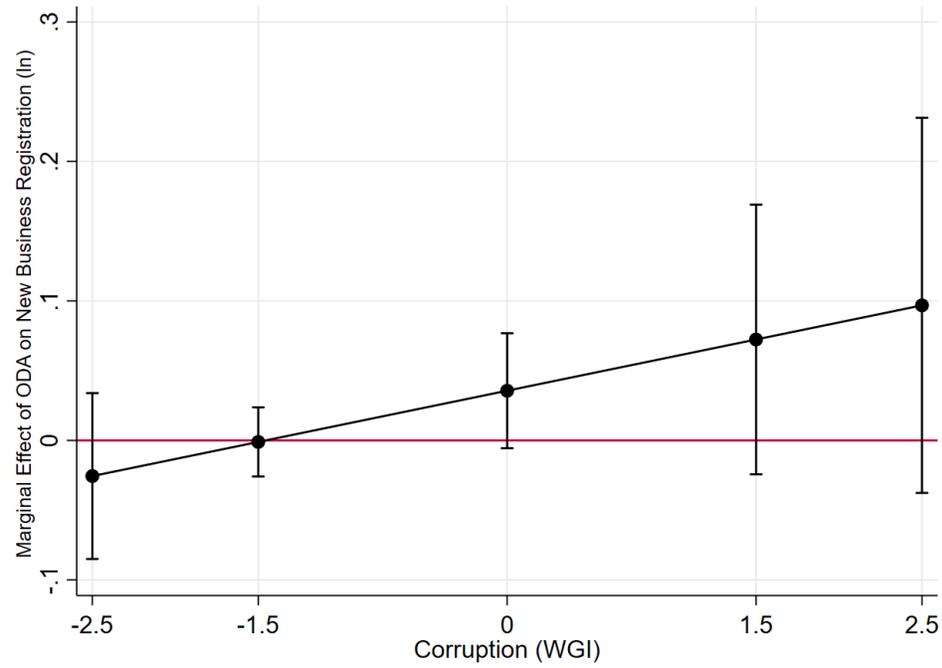

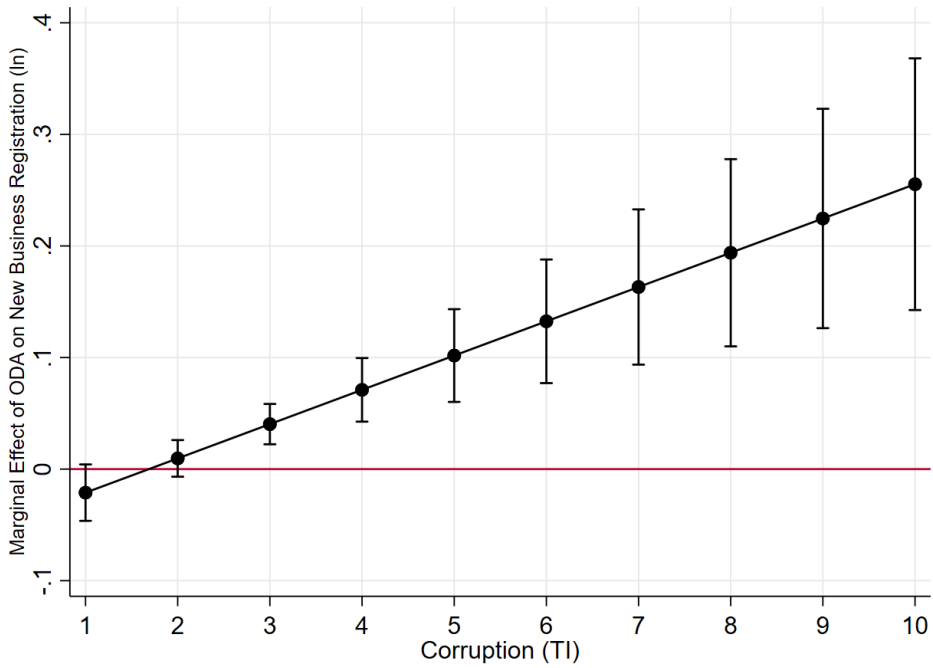

**Figure 2.** Effect of ODA on Entrepreneurship Activity, as Moderated by Corruption
*Notes.* Corruption (WGI) in panel (a) and Corruption (TI) in panel B

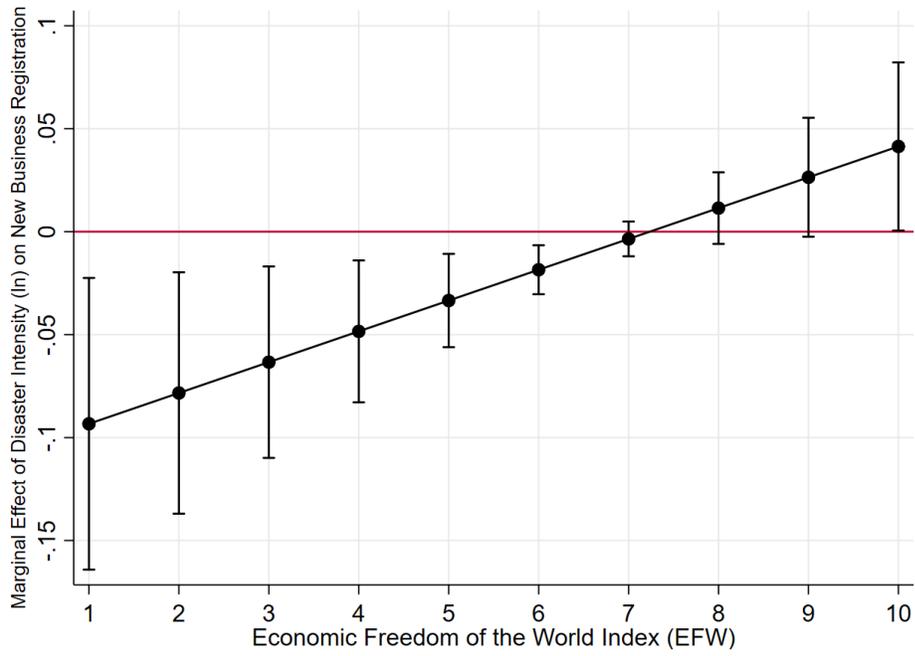
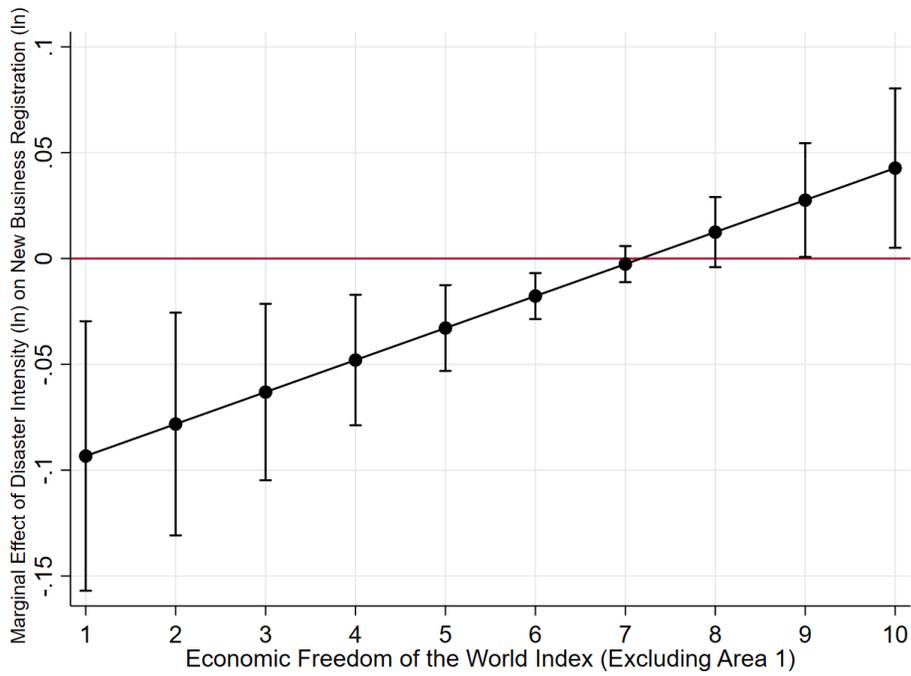

**Figure 3.** Effect of Disaster Intensity on Entrepreneurship Activity, as Moderated by Economic Freedom
*Notes.* EFW summary index in panel (a) and EFW index (excluding area 1) in panel (b).

**Table 1.** Summary Statistics

| Variable | Mean | SD | Min | Max |
|---|---|---|---|---|
| Business Start-Up Density | 1.49 | 1.9 | 0.002 | 11.88 |
| Disaster intensity (affected + injured + homeless) | 417,917 | 1,789,601 | 0 | 16,200,000 |
| Number affected by natural disaster | 407,877 | 1,778,967 | 0 | 16,100,000 |
| Injuries from natural disaster | 946 | 14,409 | 0 | 300,069 |
| Homeless due to natural disaster | 9,094 | 73,131 | 0 | 1,100,000 |
| ODA (% GNI) | 4.75 | 7.24 | 0 | 51.42 |
| GDP per capita, PPP $US | 9,397 | 7,505 | 773 | 45,636 |
| Cost of start-up procedures (% GNI) | 41.21 | 94.86 | 0.30 | 1314.6 |
| Number of days required to start a business | 27.91 | 27.05 | 2 | 224 |
| Number of procedures to register a business | 8.63 | 3.22 | 2 | 19 |
| Corruption (WGI) | -0.42 | 0.61 | -1.69 | 1.58 |
| Economic Freedom (EFW) | 7.13 | 0.76 | 3.65 | 8.89 |
| Largest population (% of Total Population) | 18.54 | 10.98 | 2 | 50 |
| GDP growth (% Annual) | 4.63 | 4.09 | -52.43 | 21.02 |
| Domestic Credit | 48.4 | 38.05 | -16.38 | 192.66 |

*Note* - N = 452 observations. * $p < 0.05$

**Table 2.** Correlation Matrix

| | | [1] | [2] | [3] | [4] | [5] | [6] | [7] | [8] | [9] | [10] | [11] | [12] | [13] | [14] | [15] |
|---|---|---|---|---|---|---|---|---|---|---|---|---|---|---|---|---|
| New registrations per 1000 people (ages 15-64) | [1] | 1 | | | | | | | | | | | | | | |
| Disaster intensity (affected + injured + homeless) | [2] | -0.38 | 1 | | | | | | | | | | | | | |
| Number affected by natural disaster | [3] | -0.35 | 0.96 | 1 | | | | | | | | | | | | |
| Injuries from natural disaster | [4] | -0.32 | 0.58 | 0.55 | 1 | | | | | | | | | | | |
| Homeless due to natural disaster | [5] | -0.24 | 0.51 | 0.39 | 0.48 | 1 | | | | | | | | | | |
| ODA (% GNI) | [6] | -0.30 | -0.07 | -0.10 | -0.09 | 0.01 | 1 | | | | | | | | | |
| GDP per capita, PPP $US | [7] | 0.61 | -0.20 | -0.16 | -0.06 | -0.17 | -0.52 | 1 | | | | | | | | |
| Cost of start-up procedures (% GNI) | [8] | -0.41 | 0.07 | 0.02 | 0.01 | 0.13 | 0.30 | -0.49 | 1 | | | | | | | |
| Number of days required to start a business | [9] | -0.17 | 0.04 | 0.04 | 0.01 | 0.05 | 0.01 | -0.16 | 0.29 | 1 | | | | | | |
| Number of procedures to register a business | [10] | -0.40 | 0.23 | 0.20 | 0.15 | 0.16 | -0.07 | -0.27 | 0.32 | 0.50 | 1 | | | | | |
| Corruption (WGI) | [11] | 0.54 | -0.24 | -0.22 | -0.13 | -0.19 | -0.09 | 0.69 | -0.34 | -0.22 | -0.44 | 1 | | | | |
| Largest population (% of Total Population) | [12] | 0.35 | -0.29 | -0.26 | -0.22 | -0.19 | -0.04 | 0.41 | -0.10 | -0.10 | -0.19 | 0.30 | 1 | | | |
| GDP growth (% Annual) | [13] | -0.17 | 0.10 | 0.09 | 0.07 | 0.12 | 0.02 | -0.13 | 0.09 | 0.10 | 0.16 | -0.17 | 0.01 | 1 | | |
| Domestic Credit | [14] | 0.44 | -0.08 | -0.04 | 0.05 | -0.08 | -0.19 | 0.54 | -0.29 | -0.24 | -0.42 | 0.65 | 0.07 | -0.21 | 1 | |
| Economic Freedom (EFW) | [15] | 0.69 | -0.17 | -0.12 | -0.10 | -0.17 | -0.25 | 0.65 | -0.40 | -0.33 | -0.49 | 0.76 | 0.35 | -0.13 | 0.59 | 1 |

*Note* - Correlations above |.04| are statistically significant, $p \leq .05$

**Table 3.** Foreign Aid Moderates the Relationship between Natural Disasters and Business Start-Up Activity

|  | Dependent variable – Business Start-Up Density (log) | | | | | |
|---|---|---|---|---|---|---|
|  | (1) | (2) | (3) | (4) | (5) | (6) |
| GDPpc (PPP) log | 1.07*** | 1.91**** | 1.98**** | 1.94**** | 1.99**** | 1.937**** |
|  | (0.298) | (0.494) | (0.500) | (0.495) | (0.499) | (0.509) |
| Cost of start-up procedures (% GNI) | -0.001* | -0.001* | -0.001* | -0.001* | -0.001 | -0.0008* |
|  | (0.0005) | (0.0004) | (0.0004) | (0.0004) | (0.0004) | (0.0004) |
| Number of days required to start a business | -0.0004 | 0.002 | 0.002 | 0.0015 | 0.0012 | 0.0017 |
|  | (0.003) | (0.004) | (0.004) | (0.004) | (0.004) | (0.004) |
| Number of procedures to register a business | -0.031 | -0.025 | -0.024 | -0.024 | -0.022 | -0.026 |
|  | (0.036) | (0.041) | (0.041) | (0.0413) | (0.042) | (0.041) |
| Corruption (WGI) | 0.249 | 0.434 | 0.454 | 0.447 | 0.457 | 0.428 |
|  | (0.195) | (0.284) | (0.283) | (0.283) | (0.284) | (0.279) |
| Population in largest city (% of Total Population) | -2.813 | -13.50** | -13.68** | -13.53** | -14.89** | -13.79** |
|  | (3.244) | (6.536) | (6.422) | (6.422) | (6.552) | (6.569) |
| GDP growth (annual %) | 0.006 | -0.001 | -0.002 | -0.002 | -0.002 | -0.001 |
|  | (0.008) | (0.009) | (0.009) | (0.009) | (0.009) | (0.009) |
| Domestic credit provided by financial sector (% of GDP) | 0.0007 | 0.0001 | -0.0002 | -0.0002 | 0.00003 | 0.0004 |
|  | (0.001) | (0.003) | (0.003) | (0.003) | (0.003) | (0.003) |
| $ODA_{t-1}$ (% GNI) |  | 0.022 | 0.017 | 0.019 | 0.017 | 0.019 |
|  |  | (0.016) | (0.015) | (0.016) | (0.014) | (0.016) |
| Natural Disaster |  |  |  |  |  |  |
| $Intensity_{t-1}$ (log) |  | -0.0013 | -0.007** |  |  |  |
|  |  | (0.003) | (0.004) |  |  |  |
| $Affected_{t-1}$ (log) |  |  |  | -0.006** |  |  |
|  |  |  |  | (0.003) |  |  |
| $Injured_{t-1}$ (log) |  |  |  |  | -0.029** |  |
|  |  |  |  |  | (0.011) |  |
| $Homeless_{t-1}$ (log) |  |  |  |  |  | -0.011*** |
|  |  |  |  |  |  | (0.004) |
| Moderating effects |  |  |  |  |  |  |
| $ODA_{t-1}$ x $Intensity_{t-1}$ (log) |  |  | 0.001** |  |  |  |
|  |  |  | (0.001) |  |  |  |
| $ODA_{t-1}$ x $Affected_{t-1}$ (log) |  |  |  | 0.0008 |  |  |
|  |  |  |  | (0.001) |  |  |
| $ODA_{t-1}$ x $Injured_{t-1}$ (log) |  |  |  |  | 0.007** |  |
|  |  |  |  |  | (0.003) |  |
| $ODA_{t-1}$ x $Homeless_{t-1}$ (log) |  |  |  |  |  | 0.0012** |
|  |  |  |  |  |  | (0.0006) |
| Constant | -9.0*** | -14.3**** | -14.8**** | -14.6**** | -14.8**** | -14.5**** |
|  | (2.88) | (3.74) | (3.80) | (3.75) | (3.72) | (3.85) |
| Number of observations | 458 | 458 | 458 | 458 | 458 | 458 |

*Note* – Dependent variable is the natural logarithm of business start-up density (NBSD). Standard errors in parentheses clustered at the country-level and robust to heteroscedasticity. Modeled using OLS with country-level and year fixed effects. * $p<0.10$, ** $p<0.05$, *** $p<0.01$, **** $p<0.001$.

**Table 4**. Regression Results

|  | DV = New Business Registrations (ln) | | | |
|---|---|---|---|---|
|  | Corruption (WGI) | Corruption (TI) | EFW | EFW No Area 1 |
|  | (1) | (2) | (3) | (4) |
| *Quality of Government* | | | | |
| ODA (% GNI) | 0.037** | 0.044*** | | |
|  | (0.018) | (0.015) | | |
| Corruption | 0.362** | -0.077 | | |
|  | (0.149) | (0.061) | | |
| ODA x Corruption | 0.019 | 0.020*** | | |
|  | (0.013) | (0.007) | | |
| *Economic Freedom* | | | | |
| Disaster Intensity (ln) | | | -0.074** | -0.068*** |
|  | | | (0.030) | (0.025) |
| Economic Freedom (EFW) | | | 0.228*** | 0.179*** |
|  | | | (0.077) | (0.067) |
| Disaster Intensity x EFW | | | 0.010** | 0.009*** |
|  | | | (0.004) | (0.003) |
| *Controls* | | | | |
| GDP per capita, PPP (ln) | 1.814**** | 2.351**** | 0.696*** | 0.729**** |
|  | (0.344) | (0.453) | (0.206) | (0.205) |
| Largest population (% total) | -11.928**** | -19.068**** | -7.71**** | -7.494**** |
|  | (2.986) | (3.664) | (2.226) | (2.224) |
| GDP growth rate | 0.005 | 0.019*** | 0.012*** | 0.012*** |
|  | (0.005) | (0.006) | (0.004) | (0.004) |
| Starting a business index | -0.119*** | -0.131*** | -0.16**** | -0.159**** |
|  | (0.043) | (0.043) | (0.029) | (0.029) |
| Domestic credit provided by financial sector | 0.001 | 0.001 | 0.001 | 0.001 |
|  | (0.002) | (0.003) | (0.001) | (0.001) |
| Constant | -14.081**** | -17.680**** | -6.67**** | -6.667**** |
|  | (2.905) | (3.854) | (1.831) | (1.825) |
| Number of observations | 470 | 362 | 682 | 682 |
| Country FE | Yes | Yes | Yes | Yes |
| Year FE | Yes | Yes | Yes | Yes |

*Note.* Robust-clustered standard errors in parentheses. Dependent variable is new business registration (log). ODA is official development assistance. WGI is World Government Institute, EFW is the Economic Freedom of the World index, PPP is the Purchasing Power Parity. GDP is Gross Domestic Product, and GNI is Gross National Income. Modeled using Ordinary Least Squares with country-level and year fixed effects. * $p<0.10$, ** $p<0.05$, *** $p<0.01$, **** $p<0.001$